## Abstract

This paper deals with sensors which compute and report linguistic assessments of their values. Such sensors, called symbolic sensors are a natural extension of smart ones when working with control systems which use artificial intelligence based technics. After having recalled the smart sensor concepts, this paper introduces the symbolic sensor ones. Links between the physical world and the symbolic one are described. It is then shown how Zadeh's approximate reasoning theory provides a smart way to implement symbolic sensors. Finally, since the symbolic sensor is a general component, a functional adaptation to the measurement context is proposed.

## Keywords

Smart Sensors, Symbolic Sensors, Intelligent Sensors

## Introduction

The concept of smart sensors or intelligent sensors introduced several years ago, is not so easy to define (Giachino [7]). One of the principal reasons is that many sensors use silicon as a transducer and have data processing integrated on the same medium (Jordan [8], Wadley [14], Muller [11]). Therefore, some authors consider these integrated sensors as intelligent ones (Middelhoeck [10], Yamasaki [15]) whereas for other authors a microprocessor based sensor should be considered as intelligent. As for us, we think that if one wishes to introduce intelligence into the sensors, it is legitimate to wonder about what kind of intelligence is to be thought of. We consider that an intelligent sensor should own four basic mechanisms: learning, faculty of reasoning, perception (the main part of the sensor) and communication. Such a sensor has not been realized yet. Each simplified implementation of these mechanisms leads to one possible "definition" of the intelligent sensor.

Nowadays, it is generally admitted that the characteristic functionality of a smart sensor should be its ability to communicate with a communication bus or network, to verify the correctness of the measurement and to adapt itself when the environment is changing (Giachino [7], Burd [3], Bois [2]). One possible comparison between an analog sensor and a smart sensor (see Fig. 1.) has been proposed by Favenec [6]. However, this comparison should be discussed since the sensor should be an analog one and be intelligent in the sense defined previously.

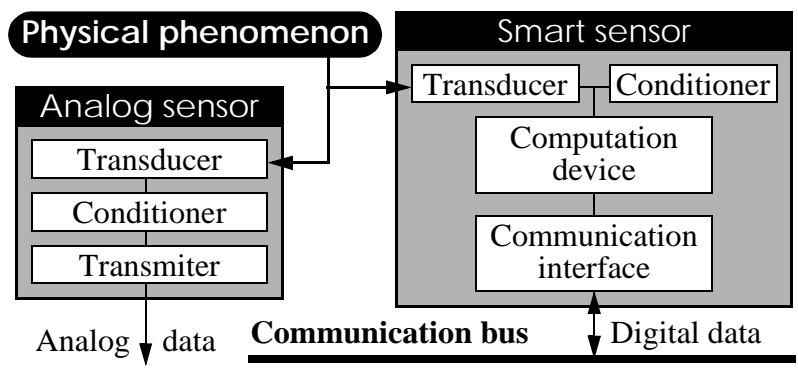

Fig. 1. Comparison between a smart sensor and an analog one

### Symbolic sensor

Nowadays, control systems become more and more complex. Artificial intelligence technics have been introduced to handle this complexity by means of new types of controllers (expert controller, intelligent controller, fuzzy controller...). As a consequence, such controllers are using more and more the symbolic coding. Rule based coding is certainly the most popular due to numbers of expert systems. For example, one can write the following rule for a temperature control problem.

      **if** temperature is high **then** reduce the control

However, even if controllers become more abstract, the perception and the control actions are still performed on the real world. Most often measurements are numerical values (e.g. the temperature in the previous example) which are converted in the controller itself. This numerical to symbolic interface can obviously be transferred into a sensor, leading to a new kind of components that we propose to call: *symbolic sensors*.

*Definition: a symbolic sensor is a smart sensor which can create and handle one or several symbolic informations relative to the measurement.*

Just as smart sensors are the logical evolution from the analogical ones due to the extensive use of microprocessors, symbolic sensors can be seen as a natural consequence of symbolic based technics. Transferring the symbolic conversion into the sensor introduces specific problems that are emphasized in the following.

### The relativity of the "symbolic measurement".

The use of a symbolic information leads us to wonder about the symbol semantics. Let us take an example in the robotics field. Assume the robot owns a distance sensor which currently returns 10cm for the distance between its tool and a workpiece. Considering an insertion task, the measurement semantics could certainly be "**the distance is long**" while in a painting case it may become "**the distance is short**". Just as for human beings, the semantics depends on the context in which the task is to be realized. Therefore, the sensor cannot be considered any longer as an independent organ, but must be seen as being part of the perception mechanism in which we include decisions and actions. From this fundamental property, one can conclude that the symbolic sensor must be configured (by a supervisor for instance) with respect to the measurement context. We exhibit here one of the major differences between a symbolic sensor and a symbolic conversion within a controller or an expert system: **the symbolic sensor is a general component that can be reconfigured to adapt itself to the measurement context**. Since symbolic sensors keep all smart sensor properties, a first functional scheme can be the following one:

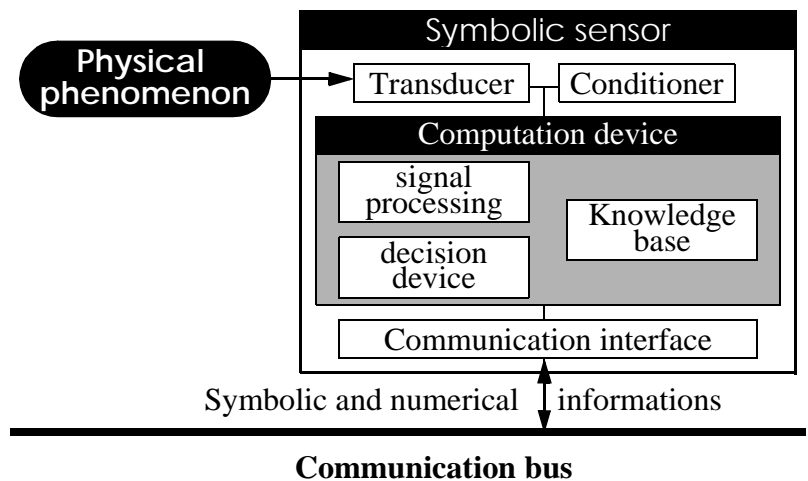

Fig. 2. Internal scheme of a symbolic sensor

The symbolic sensor concept will be emphasized in the following by a formal description of the links between the symbolic domain and the numerical one (Luzeaux [9]).

## Translation, concepts and interpretation

Let $L$ be the symbolic domain and $E$ be the numerical one. The meaning of a symbolic value will be called a translation and be defined as an injective application from the symbolic set to the set of the subsets of the numerical domain (injectivity insures that two identical symbols have the same translation).

$$\tau: L \rightarrow P(E)$$

The association of a symbolic value and its translation is called *a concept*. The symbolic measurement will be obtained by means of a new application from the numerical domain to the symbolic one, called *an interpretation*.

$$\iota: E \rightarrow L$$

Relations between the translation and the interpretation are summarized in Fig. 3.

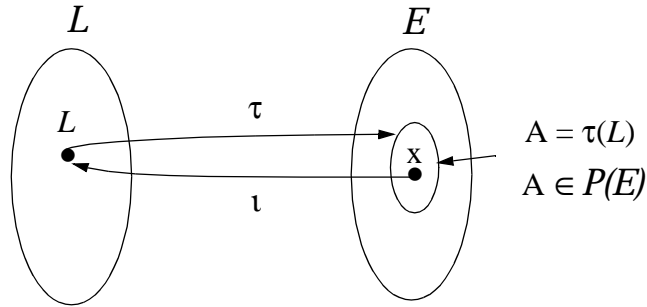

Fig. 3. Relations between a translation and an interpretation

These definitions have been used to implement symbolic sensors based on crisp sets coding (Benoit [1]). Several conditions on translations have been assumed:
- Intersections between translations are empty.
- The numerical domain is ordered with a relation noted $<_E$.
- Any translation is an interval:
  $$\forall L \in L, \ \forall x, y \in \tau(L) \text{ such that } x <_E y, \ \forall z \in E$$
  $$x <_E z <_E y \Rightarrow z \in \tau(L)$$

The numerical to symbolic interface is realized by the interpretation function. The simplest way to interpret a measurement $x$ is to return the symbol $L$ such that x belongs to its translation.

$$L = \iota(x) \text{ if } x \in \tau(L)$$

This relationship is an application when $\tau(L)$ is a partition on $E$.

## From concepts to fuzzy concepts

In the next sections, the problem of symbolic sensors dealing with imprecise or uncertain knowledge is addressed (for example, a zero detector should certainly work more around zero than at zero precisely). Zadeh's approximate reasoning theory provides a smart tool for implementing the numeric to symbolic interface. Therefore the material presented in the following is not really new in terms of symbolic manipulations based on fuzzy set (see Zadeh [17], [17]) but it provides a quite powerful way for implementing symbolic sensors. To extend the previous section to the fuzzy case, fuzzy intervals have to be defined. Let us recall that a fuzzy interval is a fuzzy subset on the set of real numbers such that (Zadeh [16])

$$\forall\ u, v,\ \forall\ x \in [u, v],\ \mu_Q(x) \geq min(\mu_Q(u), \mu_Q(v))$$

A fuzzy concept can be defined as the association of a symbol L and a fuzzy translation (i.e. a fuzzy interval). Two kinds of fuzzy intervals are usually recognized (Eshragh [5]):

- Fuzzy intervals defined as non decreasing or non increasing functions ("S" curves).
- Fuzzy intervals defined as a non increasing then non decreasing function ("Π" curves).

A general definition of fuzzy intervals by means of a parametrized class of functions (L-R functions) has been proposed by Dubois [4]. Fig. 4. presents several examples of fuzzy concepts.

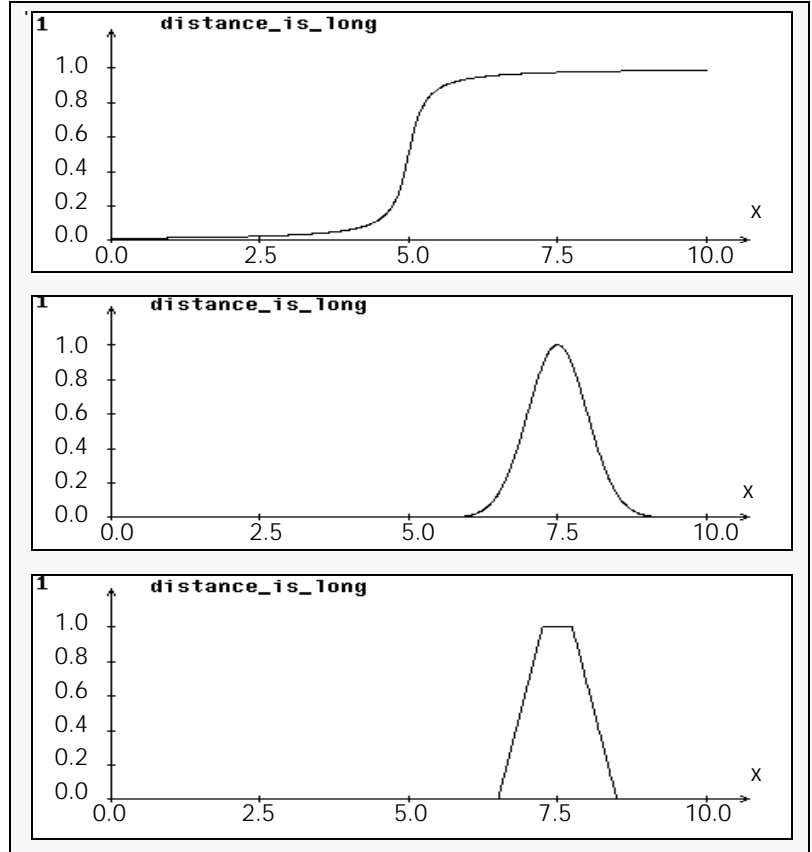

Fig. 4. Several possible definitions of the concept *distance_is_long*

### Creating new concepts

Obviously, it is a tedious task to specify each translation of the symbols. Moreover, in this case, one can say that the sensor is no more smart. The symbolic sensor works with several symbols which are in relation by means of the semantics. Let us take the example of a temperature measurement. A semantic relation links the symbolic values **hot** and **very_hot** due to the order relation on the numerical domain. This relationship between all the symbols should be managed by the sensor itself. Defining a new concept leads to give a new symbol and its translation. Modifiers, usually called linguistic hedges, can be introduced to perform such an operation. Each operation in the symbolic domain is linked to an operation on the numerical one. Let us begin with crisp sets, a function $F$ operating in the symbolic domain is defined.

$$F: L \rightarrow L$$
$$L_1 \mapsto L_2 = F(L_1)$$

Now a function $f_F$ operating on the numerical domain is in-

troduced:

$$f_F: P(E) \to P(E)$$
$$\tau(L_1) \mapsto \tau(L_2) = \tau(F(L_1)) = f_F(\tau(L_1))$$

Links between these functions are represented in Fig. 5.

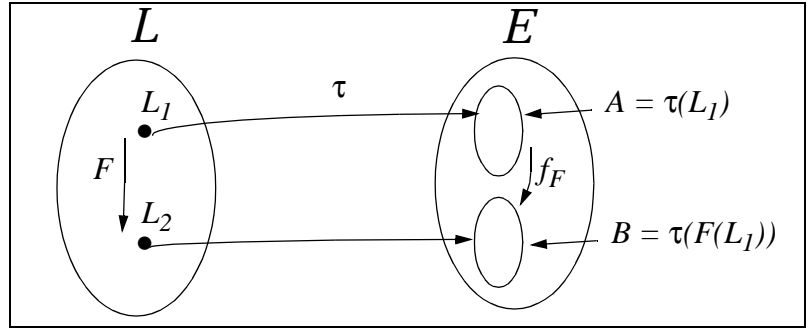

Fig. 5. Creating a new concept

The crisp case can be extended easily to the fuzzy one. The definition domain of $f_F$ becomes the set of fuzzy subsets of $E$ with the following relation on membership functions:

$$\mu_{\tau(L_2)}(x) = \mu_{\tau(F(L_1))}(x) = f_F(\mu_{\tau(L_1)}(x))$$

To simplify the reading, $\mu_{\tau(L_2)}$ will be written $\mu_{L_2}$. A more general definition has been proposed by Novak [12]:

$$\mu_{\tau(L_2)}(x) = \mu_{\tau(F(L_1))}(x) = m_F(\mu_{\tau(L_1)}(q_F(x)))$$

$m_F: [0,1] \to [0,1]$ is a *modifier* for the hedge $F$ while $q_F$ is *a translator*. Since two types of intervals have been defined, it is legitimate to define transformations between them. Four transformations can be exhibited:

- "S" curve to "S" curve
- "Π" curve to "Π" curve
- "Π" curve to "S" curve
- "S" curve to "Π" curve

Only the first three transformations seem to be of some interest for symbolic sensors: they are presented in the following.

### "S" curve to "S" curve operators.

These operators are well known, most of them have been introduced by Zadeh [17]. Let $C$ be a fuzzy concept and $\mu_C$ its associated translation, the new generated concept and its translation will be respectively noted $C1$ and $\mu_{C1}$. Several classical operators are recalled below:

| Symbolic domain | Numerical domain |
|---|---|
| $C1$ = **very** $(C)$ | $\mu_{C1}(x) = \mu^2_C(x)$ |
| $C1$ = **not**$(C)$ | $\mu_{C1}(x) = 1 - \mu_C(x)$ |
| $C1$ = **more_or_less**$(C)$ | $\mu_{C1}(x)\ \mu^{1/2}_C(x)$ |

Based on Novak 's definition, the translator is the identity for all previous cases. Let us also recall that binary operators such as $C1$ **and** $C2$, $C1$ **or** $C2$ are commonly defined by means of T-norms and T-conorms (see for example Pedrycz [13]).

A first example of a symbolic sensor can now be given. Several symbolic informations are sent to the sensor via the communication bus (see Fig. 6.). The first one, called the "**generic concept**", is used by the sensor to define all the other concepts. Here, it is defined by an arctangent function "centered" in 5m.

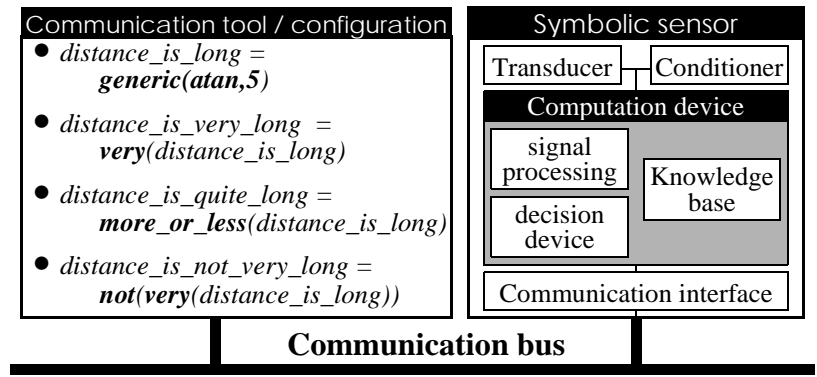

Fig. 6. Configuration of a symbolic sensor

Once these informations have been sent, the sensor is able to provide symbolic measurements. Therefore, it can return, for a given measurement, either the symbolic value (based on the interpretation function) or the grade of membership for each symbol in the lexical domain (the numerical value and many other informations such that identification, installation date, revision date... can also be obtained) (see Fig. 9.). Based on configuration informations, the sensor generates membership functions for all the measurement domain (see Fig. 7.).

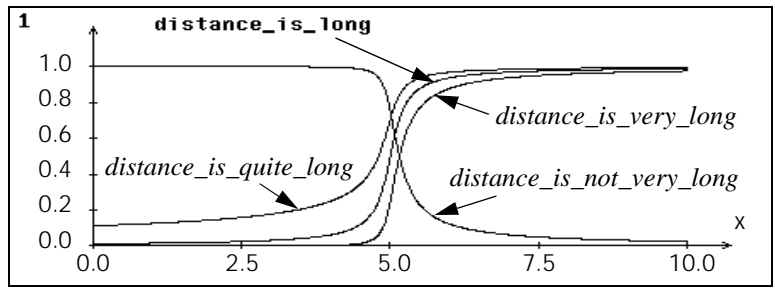

Fig. 7. New concepts generated from the generic one *distance_is_long*

## "Π" curve to "Π" curve operators

Operators defined by Zadeh act through left composition of functions. Since "Π" curves are not injective, these operators cannot be used to generate a new "Π" curve from a "Π" curve. To define such a new concept, one solution is to apply a translation. Using Novak's definition, one will now have the identity as *modifier* and $x \mapsto x \pm \Delta$ as *translator*. This operation can also be seen as the convolution of the initial concept by a delayed Dirac's distribution. Just as composition, convolution could be taken as a general mechanism to define new hedges. Using previous writing conventions, new operators can be introduced: two of them are given below ($\delta$ is the Dirac's distribution and $*$ the convolution operation).

| Symbolic domain | Numerical domain |
|---|---|
| $C1 = \textbf{more\_than}\,(C)$ | $\mu_{C1}(x) = \mu_C(x) * \delta(x-\Delta)$ |
| $C1 = \textbf{less\_than}(C)$ | $\mu_{C1}(x) = \mu_C(x) * \delta(x+\Delta)$ |

Let us give an example for a symbolic sensor dealing with "Π" curves. The generic concept is defined by a normalized gaussian function.

$$\mu_C(x) = gaussian(m_C, \sigma_C, x) = e^{-\frac{(x - m_C)^2}{2\sigma_C^2}}$$

Obviously, operators **more_than** and **less_than** give new gaussian functions defined by:

$$\mu_{\mathbf{more\_than}(C)}(x) = gaussian(m_C + \Delta, \sigma_C, x)$$
$$\mu_{\mathbf{less\_than}(C)}(x) = gaussian(m_C - \Delta, \sigma_C, x)$$

The translation value can be chosen such that the initial and the new concepts have intersecting points where the grade of membership is 0.5. For gaussian functions we have:

$$\Delta = 2 \times \sigma_C \times \sqrt{2 \times ln(2)}$$

Fig. 8. shows the example of a simple symbolic sensor dealing with three concepts defined from the generic one: *distance_is_correct*. Once again, the sensor generates the concept translations from configuration informations. The sensor can then be queried as shown in Fig. 9.

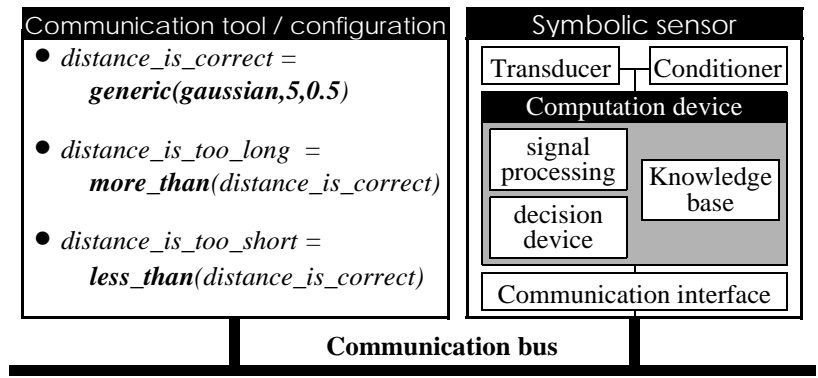

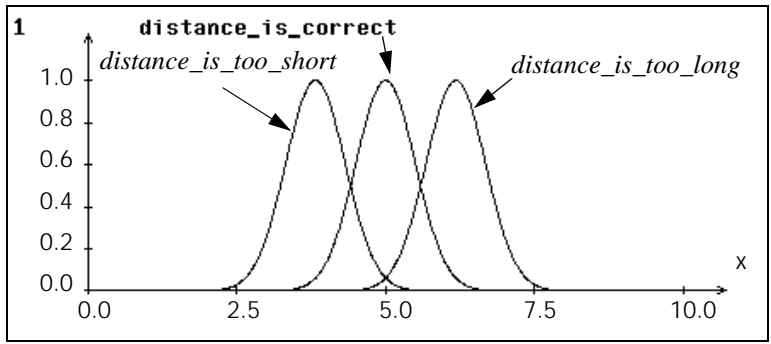

Fig. 8. Symbolic sensor based on "Π" curves

Fig. 9. Querying the sensor

## "Π" curve to "S" curve operators

To define the last type of operators, let us assume that two opposite notions exist (e.g. cold / warm, far / close, heavy / light...). They will be noted **N** and **N̄** in the following. Let *C* be a concept defined by a "Π" curve and *m* be a modal value of this concept (i.e. $\mu_C(m) = 1$). Two operators and their respective reciprocals can be defined as follows:

$\exists x_1 \geq m$ such that $\forall x > x_1$,

$\mu_{more\_N\_than(C)}(x) = \mu_{less\_\bar{N}\_than(C)}(x) > \mu_C(x)$ and
$\mu_{more\_N\_than(C)}$ is a non decreasing function

$\exists x_2 \leq m$ such that $\forall x < x_2$,

$\mu_{less\_N\_than(C)}(x) = \mu_{more\_\bar{N}\_than(C)}(x) < \mu_C(x)$ and
$\mu_{less\_N\_than(C)}$ is a non increasing function

Several definitions are possible for these operators, Eshragh and Mamdani [5] have proposed two operators **above** and **below** which verify the previous properties. Let us recall their definitions:

| Symbolic domain | Numerical domain |
|---|---|
| $C1$ = **above** $(C)$ | $\mu_{C1}(x) = 0$ if $x < m$ |
|  | $= 1 - \mu_C(x)$ else |
| $C1$ = **below**$(C)$ | $\mu_{C1}(x) = 1 - \mu_C(x)$ if $x < m$ |
|  | $= 0$ else |

Once again, let us build a simple symbolic sensor dealing with "Π" to "S" operators from a generic concept defined as a gaussian function (see Fig. 10.).

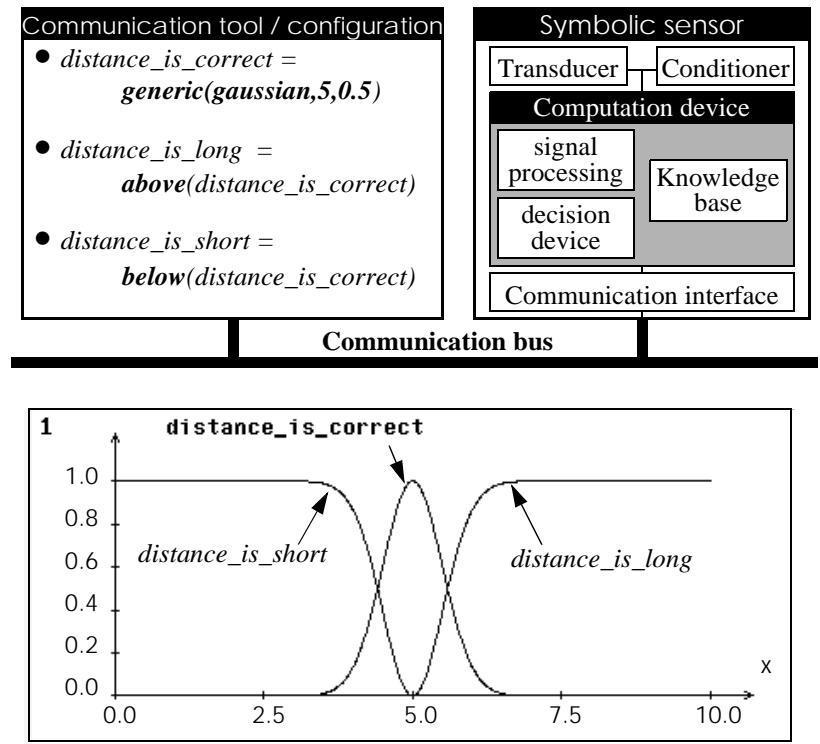

Fig. 10. Symbolic sensor using "Π" to "S" operators.

All the previous examples can be run with other generic functions (e.g. triangle, trapezoidal,....), assuming they are known by the sensor (i.e. they belong to the database). Only the generic concept parameters have to be changed in configurations.

## Taking environment into account

In the preceding, the generic concept has been taken as an a priori information given to the sensor (every other concepts is automatically worked with by means of operators). This approach preserves the semantics between concepts but does not guarantee the coherence of symbolic interpretations with respect to the measurement context (e.g. based on its configuration, the sensor returns *distance_is_very_long* while for the given context it should have returned *distance_is_long*). Therefore, taking the environment into account is nothing more than adapting the interpretation to the measurement context. Several approaches are possible, a first one based on qualitative learning with a teacher has been proposed in Benoit [1]. A functional approach which modifies interpretations through composition is presented here.

Assume the generic concept and the new generated ones are defined on the numerical domain $E$ called the **ideal universe**. Let $E'$ be the measurement domain and $h$ be a function, called the **adaptation function**, from $E'$ to $E$ (see Fig. 11.)

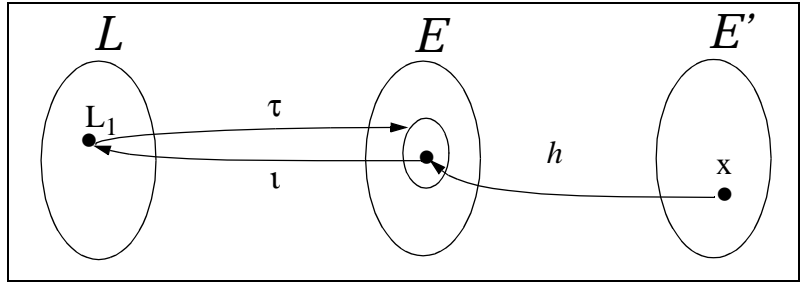

Fig. 11. Adapting the sensor to the environment

The interpretation of a measurement for crisp set is now:
$$L = \iota(h(x)) \text{ if } x \in \tau(L)$$
To run this approach, the adaptation function should be sent to the symbolic sensor. However, all adaptation functions should have specific properties that preserve the coherence of configuration informations. Especially, the generic concept definition point should not be changed by the adaptation function, furthermore the interpretation should remain linear around this point. Let $m_c$ be the generic concept definition point, the two previous properties lead to:
$$h(m_c) = m_c \text{ and } h'(m_c) = 1$$
Effects of the adaptation function on the measurement domain are called **semantic properties** (compressing or expanding the definition domain, expanding after a particular measurement...). The choice of a "good" adaptation function (i.e. having particular semantic properties and the previous mathematical properties) is not so easy. One solution is to choose an adaptation function having the desired semantics, then apply a transformation to get the mathematical properties. Let g be a function with the desired semantics at point $x_p$, then the adaptation function can be defined as follows (it is assumed that $g'(x_p) \neq 0$):
$$h(x) = a.g(x - m_c + x_p) + b$$
$$a = 1 / g'(x_p) \text{ et } b = m_c - g(x_p) / g'(x_p)$$

From these considerations, let us give four examples of functions *g* with interesting semantic properties:
- expand_before_expand_after     $g(x) = argsh(k.x)$
- compress_before_compress_after   $g(x) = sh(k.x)$
- compress_before_expand_after   $g(x) = sh(k.x)$ if $x \geq 0$
                                                $g(x) = argsh(k.x)$ else
- expand_before_compress_after   $g(x) = argsh(k.x)$ if $x \geq 0$
                                                $g(x) = sh(k.x)$ else

These functions have been chosen due to their asymptotic properties and their behavior around the origin (here $x_p = 0$). Fig. 12. shows resulting adaptation functions *h* for the generic concept definition point in 5 m. Fig. 12. presents the membership functions for a symbolic sensor dealing with seven concepts defined from a gaussian generic one whose definition point is in 5m: the adaptation function is obtained from the *g* **compress_before_expand_after** function.

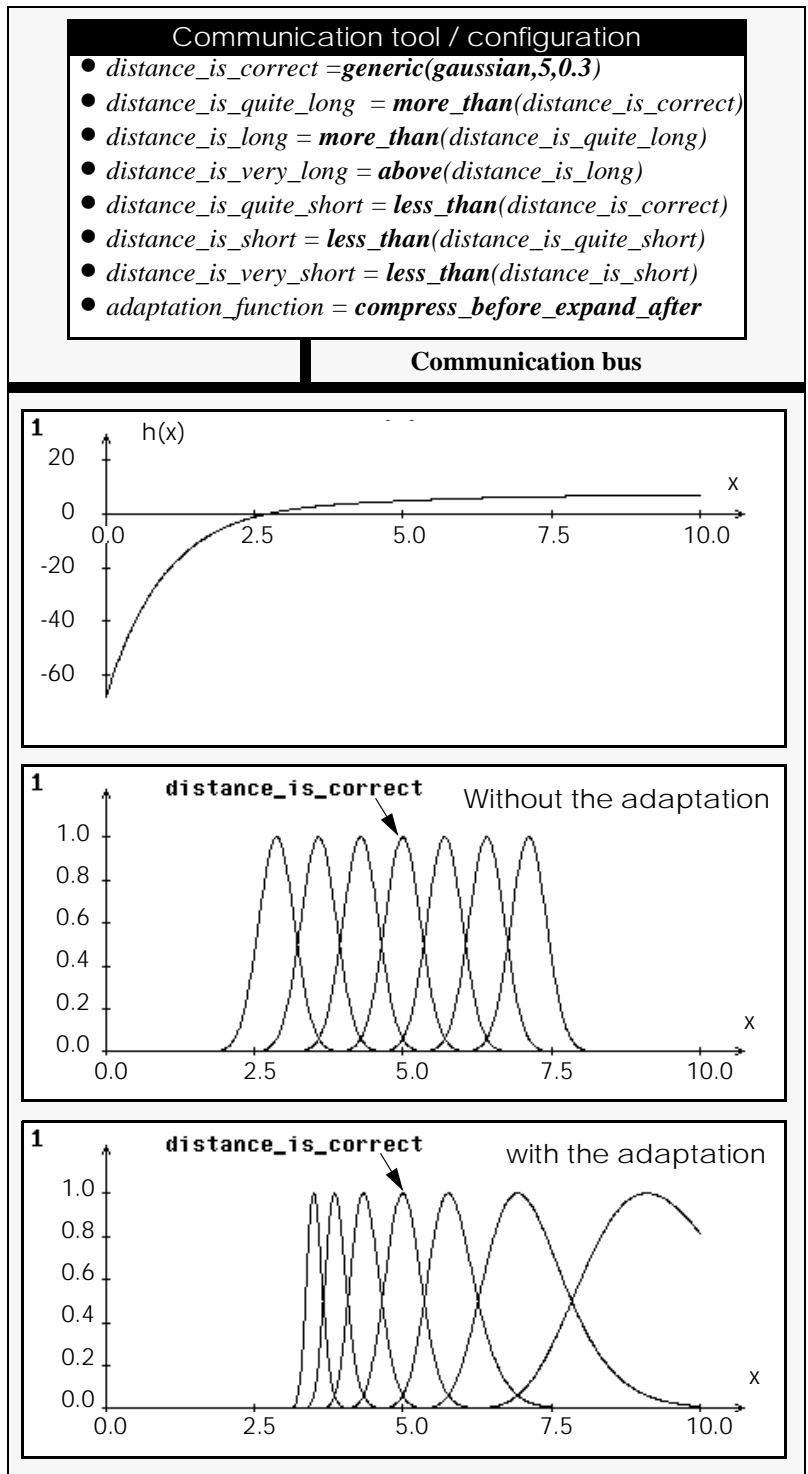

Fig. 12. Adaptating the sensor to the context

## Conclusion

Implementing a part of the knowledge and of the decision process into symbolic sensors results from the same logical evolution that lead to integrate analog to digital conversion and signal processing into smart sensors several years ago. This decentralization effort is devoted to the development of more abstract control systems that are able to work with high level informations.